\documentclass[11pt]{article}
\usepackage[square,authoryear]{natbib}
\usepackage{marsden_article}
\newcommand{\rem}[1]{}
\begin{document}

\pagestyle{myheadings}
\markright{{\it D. D. Holm and J. E. Marsden} \hfil
\underline{Measure-valued momentum map}}

\title{
Momentum Maps and Measure-valued Solutions
\\ (Peakons, Filaments and Sheets)
\\ for the EPDiff Equation}
\author{ Darryl D. Holm
\\Theoretical Division and Center for Nonlinear Studies
\\Los Alamos National
Laboratory, MS B284
\\ Los Alamos, NM 87545
\\ \footnotesize{email: dholm@lanl.gov}
\and
Mathematics Department\\
Imperial College London \\
SW7 2AZ, UK
\and
and\\
Jerrold E. Marsden
\\Control and Dynamical Systems Department, 107-81
\\California Institute of Technology
\\Pasadena, CA 91125
\\ \footnotesize{email: marsden@cds.caltech.edu}
}
\date{\footnotesize 
{\it To Alan Weinstein on the Occasion of his 60th Birthday}
\\[12pt] Started
December 2002;  This version, December 14, 2003}
\maketitle

\begin{abstract}
We study the dynamics of measure-valued solutions of what we
call the EPDiff equations, standing for the {\it Euler-Poincar\'e
equations associated with the diffeomorphism group (of
$\mathbb{R}^n$ or an
$n$-dimensional manifold $M$)}. Our main focus will
be on the case of quadratic Lagrangians; that is, on geodesic
motion on the diffeomorphism group with respect to the right
invariant Sobolev $H^1$ metric. The corresponding
Euler-Poincar\'e (EP) equations are the EPDiff equations, which
coincide with the averaged template matching equations (ATME)
from computer vision and agree with the Camassa-Holm (CH)
equations in one dimension. The corresponding equations for the
volume preserving diffeomorphism group are the well-known
LAE (Lagrangian averaged Euler) equations for
incompressible fluids.

We first show that the EPDiff equations are generated by a smooth
vector field on the diffeomorphism group for sufficiently smooth
solutions. This is analogous to known results for incompressible
fluids---both the Euler equations and the LAE
equations---and it shows that for sufficiently smooth
solutions, the equations are well-posed for short time. In
fact, numerical evidence suggests that, as time progresses,
these smooth solutions break up into singular solutions
which, at least in one dimension, exhibit soliton behavior.

With regard to these non-smooth solutions, we study measure-valued
solutions that generalize to higher dimensions the peakon
solutions of the (CH) equation in one dimension. One of the main
purposes of this paper is to show that many of the properties of
these measure-valued solutions may be understood through the
fact that their solution ansatz is a momentum map. Some
additional geometry is also pointed out, for example, that this
momentum map is one leg of a natural dual pair.
\end{abstract}

\tableofcontents

\section{Introduction}

This paper is concerned with
solutions of the EPDiff equations; that is, with the
Euler-Poincar\'e equations associated with the diffeomorphism
group in $n$-dimensions. In particular, we are concerned with
singular solutions that generalize the peakon solutions of the
Camassa-Holm (CH) equation in one dimension. The CH equation (see
\cite{CaHo1993}) for the dynamics of shallow water in a certain
asymptotic regime, is 
\begin{equation}
\label{CHEP}
u_t+3uu_x = \alpha ^2 \left( u_{xxt}+2u_x u_{xx}+uu_{xxx} \right),
\end{equation}
where $u(x,t)$ is the fluid velocity, subscripts
denote partial derivatives in position $x$ and time $t$, and
$\alpha^2$ is a positive constant. (See
\cite{DuGoHo2001,DuGoHo2003} for recent discussions of the
derivation and asymptotic validity of the CH equation for shallow
water waves, at one order beyond the Korteweg-de~Vries equation.)
Equivalently, in Hamiltonian form, the CH equation reads
\begin{equation}
\label{CHLP}
m_t+um_x + 2u_x m = 0
\end{equation}
where $m = u - \alpha ^2 u _{xx}$ and $\alpha^2 $ is a positive
constant. As \cite{CaHo1993} show, in Hamiltonian
form the CH equation is the Lie-Poisson equation associated with
the Lie algebra of one dimensional vector fields and with the
Hamiltonian
\begin{equation}
h (m) = \frac{1}{2} \int u \, m \,dx.
\end{equation}
The CH equation may be equivalently expressed in Euler-Poincar\'e
form by using the Lagrangian associated with the
$H ^1$ metric for the fluid velocity. That is, the Lagrangian  as
a function of the fluid velocity is given by the quadratic form,
\begin{equation}\label{CH-Lag}
l (u ) = \frac{1}{2} \int (u ^2 +  \alpha ^2 u _x ^2 ) \,dx.
\end{equation}

It follows from Euler-Poincar\'e theory (see \cite{MaRa1999} and
\cite{HoMaRa1998a}) that the one parameter curve of
diffeomorphisms $\eta (x_0,t)$ depending on parameter $t$ and
defined implicitly by
\[
\frac{\partial}{\partial t} \eta (x_0,t) = u ( \eta (x_0,t), t )
\]
is a geodesic in the group of diffeomorphisms of $\mathbb{R}$
(or, with periodic boundary conditions, of the circle $S ^1$)
equipped with the right invariant metric equal to the $H ^1$
metric at the identity.

A remarkable analytical property of the CH equation,
conjectured by keeping track of derivative losses in
\cite{HoMaRa1998a} and proved in \cite{Shkoller1998} is that the
geodesic equations literally define a smooth vector field in
the Sobolev $H ^s$ topology. That is, in the material
representation, the equations have no derivative loss. This
property is analogous to the corresponding results for the Euler
equations for ideal incompressible fluid flow (discovered by
\cite{EbMa1970}) and the Lagrangian averaged Euler equations
(again conjectured by \cite{HoMaRa1998a} and proved by
\cite{Shkoller1998}). 

As we will explain in \S\ref{smoothness_section}, a similar
statement holds for the $n$-dimensional EPDiff equation  if
we use the $H ^1$ metric. This is all the more
remarkable because smoothness of the geodesic flow is
{\it not true} for the $L ^2$ metric, at least not without
assuming incompressibility. Smoothness of {\em volume-preserving}
geodesic flow with respect to the $L ^2$ metric does hold for
the incompressible flow of an ideal Euler fluid, a result proved
in \cite{EbMa1970}.
\medskip

Before proceeding with a discussion of the general case of
the $n$-dimensional EPDiff equations, we shall quickly review,
mostly to establish notation, a few facts about the 
Euler-Poincar\'e and Lie-Poisson equations, whose basic
theory is explained, for example, in \cite{MaRa1999}.

\paragraph{Review of Euler-Poincar\'e and Lie-Poisson Equations.} 
Let $G$ be a Lie group and $\mathfrak{g}$ its associated Lie
algebra (identified with the tangent space to $G$ at the identity
element), with Lie bracket denoted by $[\xi, \eta]$ for $\xi, \eta
\in \mathfrak{g}$. Let $\ell: \mathfrak{g} \rightarrow \mathbb{R}$
be a given Lagrangian. Let $L: TG \rightarrow \mathbb{R} $ be the
right invariant Lagrangian on $G$ obtained by translating $\ell$
from the identity element to other points  of $G $ via the right
action of $G$ on $TG$. A basic result of Euler-Poincar\'e theory is
that the Euler--Lagrange equations for $L$ on $G$ are equivalent to
the (right) Euler-Poincar\'e equations for $\ell$ on
$\mathfrak{g}$, namely to
%-----------------------------
\begin{equation} \label{epequations}
\frac{d}{ dt} \frac{\delta \ell }{\partial \xi } =
- \operatorname{ad}_{\xi}^{\ast}\frac{\delta \ell}{\partial \xi}
\,.
\end{equation}
%-----------------------------
Here, $\operatorname{ad}_{\xi}: \mathfrak{g} \to \mathfrak{g}$ is
the adjoint operator, the linear map given by the Lie bracket
$\eta \mapsto [\xi , \eta]$. Also,
$\operatorname{ad}_{\xi}^{\ast}:
\mathfrak{g}^{\ast} \to \mathfrak{g}^{\ast}$ is its dual; that is,
$\left\langle \operatorname{ad}_{\xi}^{\ast} (\mu),  \eta
\right\rangle  = \left\langle \mu, [ \xi, \eta] \right\rangle$,
where $\left\langle \, , \right\rangle$ is the natural pairing
between $\mathfrak{g}^\ast $ and $\mathfrak{g}$. Also, $\delta \ell
/ \delta
\xi$ denotes the functional derivative of $\ell$ with respect to
$\xi
\in \mathfrak{g}$. For {\it left} invariant systems, we change the
sign in (\ref{epequations}). The Euler-Poincar\'e equations can be
written in the variational form
\begin{equation} \label{epvariational}
\delta \int \ell \,dt = 0
\,,
\end{equation}
%-----------------------------
for all variations of the form 
$\delta\xi = \dot \eta - [\xi, \eta]$
for some curve $\eta$ in $\mathfrak{g}$ that vanishes at the
endpoints.

If the reduced Legendre transformation
$\xi \mapsto \mu = \delta \ell / \delta \xi$ 
is invertible, then the Euler-Poincar\'e equations are equivalent
to the (right) Lie-Poisson equations:
\begin{equation}
\label{lpequations} 
\dot \mu  =  - \operatorname{ad} ^{* }_{\delta h/\delta \mu} \mu
\,, 
\end{equation} 
where the reduced Hamiltonian is given by,
\[
h (\mu) = \left\langle \mu, \xi\right\rangle - \ell (\xi).
\]
These equations are equivalent (via Lie-Poisson reduction and
reconstruction) to Hamilton's equations on $T ^{\ast} G $
relative to the Hamiltonian $H : T^\ast G \rightarrow
\mathbb{R}$, obtained by right translating $h$ from the identity
element to other points via the right action of $G$ on $T^\ast G$.
The Lie-Poisson equations may be written in the Poisson bracket
form
\begin{equation} \label{lppoisson}
\dot{F} = \left\{ F, h \right\},
\end{equation}
where $F: \mathfrak{g}^\ast \rightarrow \mathbb{R}$ is
an arbitrary smooth function and the bracket is the
(right) Lie-Poisson bracket given by
\begin{equation}
\label{lpb} \{F, G\}(\mu )  
=  \left\langle \mu , \left[ \frac{ \delta F}{\delta  \mu},
\frac{\delta  G}{\delta \mu } \right] \right\rangle . 
\end{equation} 

In the important case when $\ell$ is quadratic, the Lagrangian $L $
is  the quadratic form associated to a right invariant Riemannian
metric on $G$. In this case, the Euler--Lagrange equations for $L$
on $G$ describe geodesic motion relative to this metric and these
geodesics are then equivalently described by either the
Euler-Poincar\'e, or the Lie-Poisson equations. 

\paragraph{Outline of the paper.}
The main results of the present paper are as follows:
\begin{enumerate}
\item In \S\ref{epdiff_section} we review some basic facts about
the EPDiff equations, especially the {\it singular solution ansatz}
\textup{(\ref{m-ansatz-intro})} of \cite{HoSt2003}
that introduces a class of singular solutions that
generalize the peakon solutions of the CH equation to higher
spatial dimensions. 
\item In \S\ref{smoothness_section} we give a plausibility
argument that the EPDiff equations possess an interesting {\it
smoothness property}; namely, they define a smooth vector field
(that is, they define ODE's with no derivative loss) in the
Lagrangian representation. This means, in particular, that the
EPDiff equations are locally well posed for sufficiently smooth
initial data. Because of the development of singularities in
finite time, which the numerics suggests, the smooth solutions
may not exist globally in time. This smoothness property is
similar to the corresponding smoothness property of the Euler
equations for ideal incompressible fluid mechanics shown in
\cite{EbMa1970}.
\item In \S\ref{SSMomMap_section}, we show that the singular
solution ansatz \textup{(\ref{m-ansatz-intro})} defines an
equivariant momentum map. We do this in a natural way by
identifying the singular solutions with certain curves in the
space of embeddings $\operatorname{Emb}(S,
\mathbb{R}^n)$ of a generally lower dimensional manifold $S$ into
the underlying manifold $\mathbb{R}^n$ (or an $n$-manifold $M$)
and letting the diffeomorphism group act on this space. The right
action of $\operatorname{Diff} (S)$ corresponds to the right
invariance of the EPDiff equations, while the left action of
$\operatorname{Diff}(\mathbb{R}^n)$ gives the desired solution
ansatz. 
\item In \S\ref{geom_mommap} we explore the geometry of the
singular solution momentum map, in parallel with the
corresponding work on singular solutions (vortices, filaments,
etc.) for the Euler equations of an ideal fluid in 
\cite{MaWe1983}.
\item Finally, in \S\ref{future_section}, we discuss some of the
remaining challenges and speculate on some of the many possible
future directions for this work. 
\end{enumerate}

\paragraph{Historical Note.} This paper is dedicated to our friend
and collaborator Alan Weinstein and, for us, this work parallels
some of  our earlier collaborations with him. Alan's basic works
on reduction, Poisson geometry, semidirect product theory, and
stability in mechanics---just to name a few areas---have been,
and remain incredibly influential and important to the field of
geometric mechanics. See, for instance, \cite{MaWe1974,
Weinstein1983b, MaRaWe1984, Weinstein1984, HoMaRaWe1985} 
\medskip

\rem{
In a general sea of confusion in the period up to the early 1980's
concerning mechanics on Lie groups, coadjoint orbits, brackets, and
so on, it was Alan's insight, being able to see further and
clearer than others, together with his historical researches (see
\cite{Weinstein1983a}) that not only led him to first rate
scientific discoveries, but it also led \cite{MaWe1983} to name
the Lie-Poisson equations as such, and this name, being very
appropriate, has stuck. This is just one of many examples of his
impact and influence on the area.}

\medskip

Mechanics on Lie groups was pioneered by \cite{Arnold1966}, a
reference that is a key foundation for the subject and in
particular for the present paper. However, this theory was in a
relatively primitive state, even by 1980, and it has benefited
greatly from Alan's insights. In fact, the clear distinction
between the Euler-Poincar\'e and Lie-Poisson equations, the
former with a variational structure and the latter with its
Poisson structure, which took until the 1980's to crystalize,
was greatly aided by Alan's work.
\medskip

Alan has made key contributions to many fundamental concepts in
geometrical mechanics, such as Lagrangian submanifolds and
related structures (\cite{Weinstein1971, Weinstein1977}),
symplectic reduction (\cite{MaWe1974}), normal modes and
periodic orbits (\cite{Weinstein1973, Weinstein1978}), Poisson
manifolds (\cite{Weinstein1983b}), geometric phases
(\cite{Weinstein1990}), Dirac structures (\cite{CoWe1988})
groupoids and Lagrangian reduction (\cite{Weinstein1996}) and
the plethora of related ``oid'' structures he has been working
on during the last decade (just look over the over 200 papers on
MathSciNet he has written!) that will surely play an important
role in the next generation of people working in the area of
geometric mechanics.

Of Alan's papers, the one that is most directly relevant to the
topics discussed in the present paper is \cite{MaWe1983}.  Alan
himself is is still developing this area, too, as in
\cite{Weinstein2002}.

\section{The EPDiff Equation} \label{epdiff_section} In this
section we review the EPDiff equation; that is, the
Euler-Poincar\'e (EP) equation associated with the diffeomorphism
group in $n$-dimensions. This equation coincides with a limiting
case of the CH equation for shallow water waves in one and two
dimensions. It also coincides with the ATME equation (the averaged
template matching equation) in two dimensions. The latter equation
arises in computer vision; see, for instance, \cite{Mumford1998,
HiMaAr2001} or \cite{MiTrYo2002} for a description and further
references. We have chosen to call this by a generic name, the
{\bfi EPDiff equation}, because it has these various 
interpretations in different applications. Of course these
different interpretations also provide opportunities: for
example, to see to what extent the singular solutions found in
the EPDiff equations are applicable, either for shallow water
wave interactions, or for computer vision applications. A recent
combination of these ideas in which image processing imitates
soliton interactions appears in
\cite{HoTrYo2003}.

\paragraph{Statement of the EPDiff Equations.} Treating analytical
issues formally at this point, let
$\mathfrak{X}$ denote the Lie algebra of vector fields on an
$n$-dimensional manifold $M$ (such as $\mathbb{R}^n$). The
vector fields comprise the algebra associated with the
diffeomorphism group of $M$, but the usual Jacobi-Lie bracket is
the negative of the (standard) Lie algebra bracket.  (See
\cite{MaRa1999} for a discussion.)

Let $\ell:
\mathfrak{X} \rightarrow \mathbb{R}$  be a given  Lagrangian and
let $\mathfrak{M}$ denote the space of one-form densities on $M$,
that is, the momentum densities. The corresponding momentum
density of the fluid is defined as
\[
m = \frac{\delta \ell }{\delta u } \in \mathfrak{M}
\,,
\]
which is the functional derivative of the Lagrangian $\ell$ with
respect to the fluid velocity $u \in \mathfrak{X}$. If $u$ is
the basic dynamical variable, the EPDiff equations are simply
the Euler-Poincar\'e equations  associated with this Lagrangian.
Equivalently, if $m$ is the basic dynamical variable, a
Legendre transformation allows one to identify the EPDiff
equations as the Lie-Poisson equations associated with the
resulting Hamiltonian. For the case of $\mathbb{R}^n$, we will
use vector notation for the momentum density
$\mathbf{m}(\mathbf{x},t):\,
\mathbb{R}^n\times{\mathbb{R}}\to{\mathbb{R}^n}$ (a bold
$\mathbf{m}$ instead of a lightface $m$). The EPDiff equations
are as follows (see
\cite{HoMaRa1998a,HoMaRa1998b, HoMaRa2002} for
additional background and for techniques for computing the
Euler-Poincar\'e equations for field theories),
\begin{equation}\label{EP-eqn-vec-intro}
\frac{\partial }{\partial t}\mathbf{m} \
+
\underbrace{\
\mathbf{u}\cdot\nabla \mathbf{m}\
}_{\hbox{convection}}\
+\
\underbrace{\
\nabla \mathbf{u}^T\cdot\mathbf{m}\
}_{\hbox{stretching}}\
+\
\underbrace{\
  \mathbf{m}\,(\operatorname{div}\mathbf{u})\
}_{\hbox{expansion}}
=0
\,.
\end{equation}

In coordinates $x ^i$, $i=1,2,\dots,n$, using the summation
convention, and writing $\mathbf{m} = m _i d x ^i \otimes d ^n x
$ (regarding $\mathbf{m}$ as a one-form density) and $\mathbf{u}
= u ^i \partial / \partial x ^i$ (regarding $\mathbf{u}$ as a
vector field), the EPDiff equations read
\begin{equation} \label{epdiffcoord}
\frac{\partial}{\partial t} m _i
+ u ^j \frac{\partial m_i}{\partial x^j}
+ m _j \frac{\partial u ^j }{\partial x ^i }
+ m _i \frac{\partial u ^j }{\partial x ^j}
=0
\,.
\end{equation}

The EPDiff equations can also be written very nicely as

\begin{equation} \label{epdifflie}
\frac{\partial \mathbf{m} }{ \partial t } + \pounds _{\mathbf{u}}
\mathbf{m} = 0,
\end{equation}
where $\pounds _{\mathbf{u}} \mathbf{m}$ denotes the Lie
derivative of the momentum one form density $\mathbf{m}$ with
respect to the velocity vector field $\mathbf{u}$. 

As mentioned earlier, if $\ell$ is a quadratic
function of $\mathbf{u}$, then the EPDiff equation
((\ref{EP-eqn-vec-intro}) or, equivalently (\ref{epdifflie})),
is the Eulerian description of geodesic motion on the
diffeomorphism group of the underlying space (in this case
$\mathbb{R}^n$). The corresponding metric is the right invariant
metric on the group whose value on the Lie algebra (the group's
tangent space at the identity---the space of vector fields),
is defined by $\ell$. Since the Lagrangian $\ell$ is
positive and quadratic in $\mathbf{u}$, the momentum density is
linear in $\mathbf{u}$ and so defines a positive symmetric
operator
$Q_{\rm op}$ by
\[
\mathbf{m} = \frac{\delta  \ell}{\delta 
\mathbf{u}} = Q_{\rm op}\mathbf{u}\,.
\]

\paragraph{Variational Formulation.} Following the variational
formulation of EP theory, the particular EP equation
(\ref{EP-eqn-vec-intro}) may be derived from the following
constrained variational principle:
\[
\delta \int \ell ( \mathbf{u} ) \,dt = 0
\,.
\]
The variations are constrained to have the form
\[
\delta \mathbf{u} = \dot{\mathbf{w}} 
+  \mathbf{w} \cdot \nabla \mathbf{u} -
\mathbf{u} \cdot \nabla
\mathbf{w}.
\] 
This assertion may of course be verified directly. These
constraints are analogous to the so-called ``Lin constraints''
used for a similar variational principle for fluid mechanics.
(See \cite{MaRa1999} for a discussion and references.)

\paragraph{Hamiltonian Formulation.} 
A Legendre transformation yields the Hamiltonian,
\[
H(\mathbf{m})
= \left\langle \mathbf{m}, \mathbf{u}
\right\rangle -\ell(\mathbf{u})
\,,
\]
where $\langle\,,\,\rangle$ is the natural pairing between
one form densities and vector fields given by integration.
This Hamiltonian is the corresponding quadratic form for the
momentum,
\begin{equation}\label{EP-momentum-norm}
H(\mathbf{m})
=
 \ell ( Q_{\rm op}^{-1} (\mathbf{m}) )
\,.
\end{equation}
Of course it often happens that $Q_{\rm op}$ is
a differential operator and in this case the
inverse is usually given in terms of
the convolution with the Green's function $G$, corresponding to the
appropriate solution domain and boundary conditions;
\[
\mathbf{u}
= 
\frac{\delta H(\mathbf{m})}{\delta\mathbf{m}}
=
G*\mathbf{m}
\,.
\]

According to the general theory, the EP equation
(\ref{EP-eqn-vec-intro}) may be expressed in Hamiltonian form
by using the Lie-Poisson bracket on $\mathfrak{M}$ as
\begin{equation}\label{LP-eqn}
\frac{\partial }{\partial t}\mathbf{m}
=
\{\mathbf{m}\,,\,H\}_{LP}
=-
{\rm\,ad\,}^*_{\delta{H}/\delta\mathbf{m}}\mathbf{m}
\,.
\end{equation}

\paragraph{One-dimensional CH Peakon Solutions.}
We return now to the CH equation (\ref{CHLP}), which, as  we have
noted, is the same as the EPDiff equation
(\ref{EP-eqn-vec-intro}) for the case of one spatial dimension,
when the momentum velocity relationship is defined by the Helmholtz
equation, $m = u -\alpha^2 u_{xx}$. In one dimension, the CH equation
has solutions whose momentum is supported at points on the real line
via the following sum over Dirac delta measures,
\begin{equation}\label{m-peakon-ansatz-intro}
m(x,t)=\sum_{i=1}^Np_i(t)\,\delta\big(x-q_i(t)\big)
\,.
\end{equation}
The velocity corresponding to this measure-valued momentum is
obtained by convolution with the Green's function,
$$G(|x-y|)=\tfrac{1}{2}e^{-|x-y|/\alpha}\,,$$
for the one-dimensional Helmholtz operator,
$Q_{\rm op}=(1-\alpha^2\partial_x^2)$, appearing in the CH momentum
velocity relationship, $m = Q_{\rm op}u$.
Consequently, the CH velocity corresponding to this momentum is given
by a superposition  of peaked traveling wave pulses,
\begin{equation}\label{u-peakon-ansatz-intro}
u(x,t)=\frac{1}{2}\sum_{i=1}^Np_i(t)\,e^{-|x-q_i(t)|/\alpha}
\,.
\end{equation}
Thus, the superposition of ``peakons'' in velocity arises from the
delta function solution ansatz (\ref{m-peakon-ansatz-intro}) for
the momentum. 
\medskip

\begin{figure}[ht]
\begin{center}
\includegraphics[scale=0.5,angle=0]{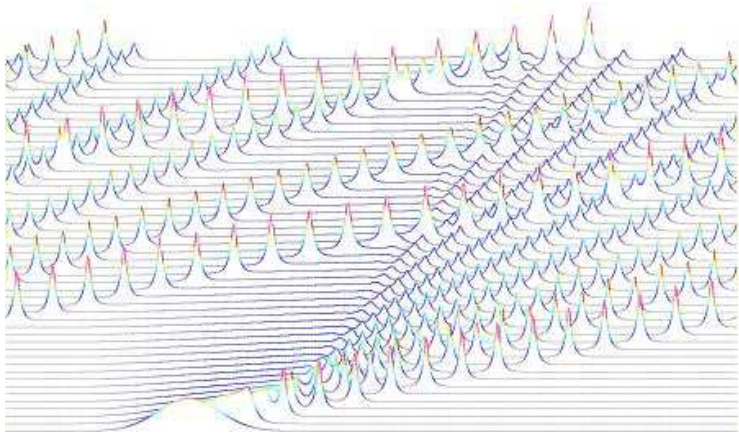}
\end{center}
\caption{\footnotesize
This figure shows a smooth localized
(Gaussian)  initial condition for the CH equation breaking up into
an ordered train of peakons as time evolves (the time direction
being vertical, which then eventually wrap around the periodic
domain and interacting with other slower emergent peakons and
causing a phaseshift (c.f. \cite{AlMa1992}).}
\label{peakon_figure}
\end{figure}

Remarkably, the isospectral eigenvalue problem
for the CH equation implies that {\it only} these singular solutions
emerge asymptotically in the solution of the initial value problem
in one dimension, \cite{CaHo1993}. Figure \ref{peakon_figure}
shows the emergence of peakons from an initially Gaussian
velocity distribution and their subsequent elastic collisions in
a periodic one-dimensional domain.\footnote{Figure
\ref{peakon_figure} was kindly supplied by Martin Staley} This
figure demonstrates that singular solutions dominate the initial
value problem and, thus, that it is imperative to go beyond
smooth solutions for the CH equation; the situation is similar
for the EPDiff equation.

Remarkably, the dynamical equations for $p_i(t)$ and $q_i(t)$,
$i=1,\dots,N$, that arise from solution ansatz
(\ref{m-peakon-ansatz-intro}-\ref{u-peakon-ansatz-intro})
comprise an integrable system for any $N$. This system is
studied  in (\cite{AlCaFeHoMa2001}) and references therein.  See
also \cite{Va2002,Va2003} for discussions of how the
integrable dynamical system for $N$ peakons is related to the
Toda chain with open ends.

\paragraph{Generalizing the CH peakon solutions to $n$ dimensions}
Building on the peakon solutions for the CH equation and the
pulsons for its generalization to other traveling-wave shapes
(see \cite{FrHo2001}), \cite{HoSt2003} introduced the following
measure-valued (that is, density valued) ansatz for the
$n-$dimensional solutions of the EPDiff
equation (\ref{EP-eqn-vec-intro}):
\begin{equation}\label{m-ansatz-intro}
\mathbf{m}(\mathbf{x},t)
=
\sum_{a=1}^N\int\mathbf{P}^a(s,t)\,
\delta\big(\,\mathbf{x}-\mathbf{Q}^a(s,t)\,\big)ds.
\end{equation}
These solutions are vector-valued 
functions supported in $\mathbb{R}^n$
on a set of $N$ surfaces (or curves) of  codimension $(n-k)$ for
$s\in \mathbb{R}^{k}$ with $k<n$.  They may, for example, be supported on
sets of points (vector peakons, $k=0$), one-dimensional filaments
(strings, $k=1$), or two-dimensional surfaces (sheets, $k=2$) in
three dimensions. 

{\it One of the main results of this paper is that the
singular solution ansatz \textup{(\ref{m-ansatz-intro})} is a
momentum map. This result helps to organize the theory and to
suggest new avenues of exploration, as we shall explain.}

 Substitution of the solution ansatz
(\ref{m-ansatz-intro}) into the EPDiff
equations (\ref{EP-eqn-vec-intro}) implies the following 
integro-partial-differential equations (IPDEs) for the evolution of
such strings and sheets,
%%%%%%%%%%%%%%%%%%%%%%%%%%%%%%%%%%%%%%%%%%%%%%%%%%%%%%%%%%%%%%%%%%%%
\begin{eqnarray}
%\hspace{-3mm}
\frac{\partial }{\partial t}\mathbf{{Q}}^a (s,t)
\!\!&=&\!\!
\!\!\sum_{b=1}^{N} \int\mathbf{P}^b(s^{\prime},t)\,
G(\mathbf{Q}^a(s,t)-\mathbf{Q}^b(s^{\prime},t)\,\big)ds^{\prime}
\,,\label{IntDiffEqn-Q}\\
%\hspace{-3mm}
\frac{\partial }{\partial t}\mathbf{{P}}^a (s,t)
\!\!&=&\!\!
-\,\!\!\sum_{b=1}^{N} \int
\big(\mathbf{P}^a(s,t)\!\cdot\!\mathbf{P}^b(s^{\prime},t)\big)
\, \frac{\partial }{\partial \mathbf{Q}^a(s,t)}
G\big(\mathbf{Q}^a(s,t)-\mathbf{Q}^b(s^{\prime},t)\big)\,ds^{\prime}
\,.
\nonumber
\end{eqnarray}
%%%%%%%%%%%%%%%%%%%%%%%%%%%%%%%%%%%%%%%%%%%%%%%%%%%%%%%%%%%%%%%%%%%%%%%%
%
Importantly for the interpretation of these solutions, the coordinates
$s\in \mathbb{R}^{k}$ turn out to be Lagrangian coordinates. The 
velocity field corresponding to the momentum solution ansatz
(\ref{m-ansatz-intro}) is given by
\begin{equation}\label{u-ansatz-intro}
\mathbf{u}(\mathbf{x},t)
=
G*\mathbf{m}
=
\sum_{b=1}^N\int\mathbf{P}^b(s^{\prime},t)\,
G\big(\,\mathbf{x}-\mathbf{Q}^b(s^{\prime},t)\,\big)ds^{\prime}
\,,\quad
\mathbf{u}\in{\mathbb{R}^n}
\,.
\end{equation}
When evaluated along the curve $\mathbf{x}=\mathbf{Q}^a(s,t)$, the
velocity satisfies,
\begin{equation}\label{Qdot-ansatz-intro}
\mathbf{u}(\mathbf{x},t)\Big|_{\mathbf{x}=\mathbf{Q}^a(s,t)}
=
\sum_{b=1}^N\int\mathbf{P}^b(s^{\prime},t)\,
G\big(\,\mathbf{Q}^a(s,t)
-\mathbf{Q}^b(s^{\prime},t)\,\big)ds^{\prime}
=
\frac{\partial\mathbf{Q}^a(s,t)}{\partial t}
\,.
\end{equation}
Thus, the lower-dimensional support sets defined on
$\mathbf{x}=\mathbf{Q}^a(s,t)$ and parameterized by coordinates
$s\in{\mathbb{R}}^{k}$ move with the fluid velocity. Moreover, equations
(\ref{IntDiffEqn-Q}) for the evolution of these support sets
are canonical Hamiltonian equations,
%-----------------------------
\begin{equation}
\label{IntDiffEqns-Ham}
\frac{\partial }{\partial t}\mathbf{{Q}}^a (s,t)
=
\frac{\delta H_N}{\delta \mathbf{P}^a}
\,,\qquad
\frac{\partial }{\partial t}\mathbf{{P}}^a (s,t)
=
-\,\frac{\delta H_N}{\delta \mathbf{Q}^a}
\,.
\end{equation}
%-----------------------------
The Hamiltonian function 
$H_N:(\mathbb{R}^n\times \mathbb{R}^n)^{N}\to \mathbb{R}$ is,
%-----------------------------
\begin{equation} \label{H_N-def}
H_N = \frac{1}{2}\!\int\!\!\!\!\int\!\!\sum_{a\,,\,b=1}^{N}
\big(\mathbf{P}^a(s,t)\cdot\mathbf{P}^b(s^{\prime},t)\big)
\,G\big(\mathbf{Q}^a(s,t)-\mathbf{Q}^b(s^{\prime},t)\big)
\,ds\,ds^{\prime}
\,.
\end{equation}
%-----------------------------
This is the Hamiltonian for geodesic motion on the cotangent bundle of
a set of curves $\mathbf{Q}^a(s,t)$ with respect to the metric given by
$G$. This dynamics was investigated numerically in 
\cite{HoSt2003} to which we refer for more details of
the solution properties.

One of our main goals is to show that the solution ansatz
(\ref{m-ansatz-intro}) can be phrased in terms of a momentum 
map that naturally arises in this problem. This geometric
feature underlies the remarkable reduction properties of the
EPDiff equation and ``explains'' why the preceding
equations must be Hamiltonian, namely, because momentum maps are
Poisson maps. 

As explained in general terms in \cite{MaWe1983}, the way one
implements a coadjoint orbit reduction is through a momentum
map, and this holds even for the case of singular orbits (again
ignoring functional analytic details). Thus, in summary, {\it
the reduction
\textup{(\ref{IntDiffEqns-Ham})} is the {\rm EPDiff} analog of the
reduction in fluid mechanics (that is, the
\textup{EPDiff$_{\rm Vol}$} equations) to point (or blob) vortex
dynamics, vortex filaments, or sheets.}

There are, however, some important differences between vortex 
dynamics for incompressible flows and the dynamics of the measure
valued EPDiff solutions. For example, the Lagrangian representations
of the equations of motion show that EPDiff solutions have
inertia, while the corresponding solutions for point (or blob)
vortices of the EPDiff$_{\rm Vol}$ dynamics have no inertia. That is,
the equations of motion for measure valued solutions on EPDiff$_{\rm
Vol}$ are {\it first order} in time, while the dynamical equations
for measure valued solutions on EPDiff are {\it second order} in time.
This difference has profound effects on the properties of the
solutions, especially on their stability properties. Numerical
investigations of \cite{HoSt2003} show, for example, that the
codimension-one solutions  of EPDiff are stable, while higher
codimension solutions of EPDiff are very unstable to
codimension-one perturbations. In contrast, the codimension-two
solutions  of EPDiff$_{\rm Vol}$ are known to be stable.

\paragraph{Comments on the Physical Meaning of the Equations.}
The EPDiff equations with the Helmholtz relation between velocity
and momentum are not quite the CH equations for surface waves in
2D. Those would take precisely the same form, but the shallow
water wave relation in the 2D CH approximation would be 
\[
      m = u - \alpha^2 \operatorname{Grad} \operatorname{Div} u;
\quad \mbox{that is,} \quad m_i = u_i - \alpha^2 u_{j,ji}
\] 
rather than the Helmholtz operator form,
\[
      m = u - \alpha^2 \operatorname{Div} \operatorname{Grad} u 
\quad \mbox{that is,} \quad 
m_i = u_i - \alpha^2 u_{i,jj}
\]
The corresponding Lagrangians are, respectively,
\begin{equation}\label{Div-norm}
l_{\rm CH} (u ) = \frac{1}{2} \int (|u|^2 +  \alpha ^2
(\operatorname{Div} u) ^2 ) 
\,dx\,dy.
\end{equation}
and
\begin{equation}\label{Grad-norm}
l_{\rm EPDiff} (u ) = \frac{1}{2} \int (|u|^2 +  \alpha ^2
|\operatorname{Grad} u| ^2 ) 
\,dx\,dy.
\end{equation}

This difference was noted in \cite{KrScDu2001}, which
identified (\ref{Div-norm}) as the generalization of
(\ref{CH-Lag}) for water waves in two dimensions.
One may also verify this by considering the limit of the Green-Nagdhi
equations for small  potential energy. (The CH equation arises
in this limit. The Lake and Great Lake equations of
\cite{CaHoLe1996,CaHoLe1997} also arise in a variant
of this limit.)

Remarkably, the numerics in \cite{HoSt2003} show that the
solutions for a variety of initial conditions are
indistinguishable in these two cases. 
The initial conditions in \cite{HoSt2003} were all spatially
confined velocity distributions.

Notice that this difference affects the choice of Hamiltonian, but
the equations are still Euler-Poincar\'e equations for the
diffeomorphism group and the description of the anzatz
(\ref{m-ansatz-intro}) as a momentum map is independent of this
difference in the equations.

\begin{figure}[ht]
\begin{center}
\includegraphics[scale=0.7,angle=0]{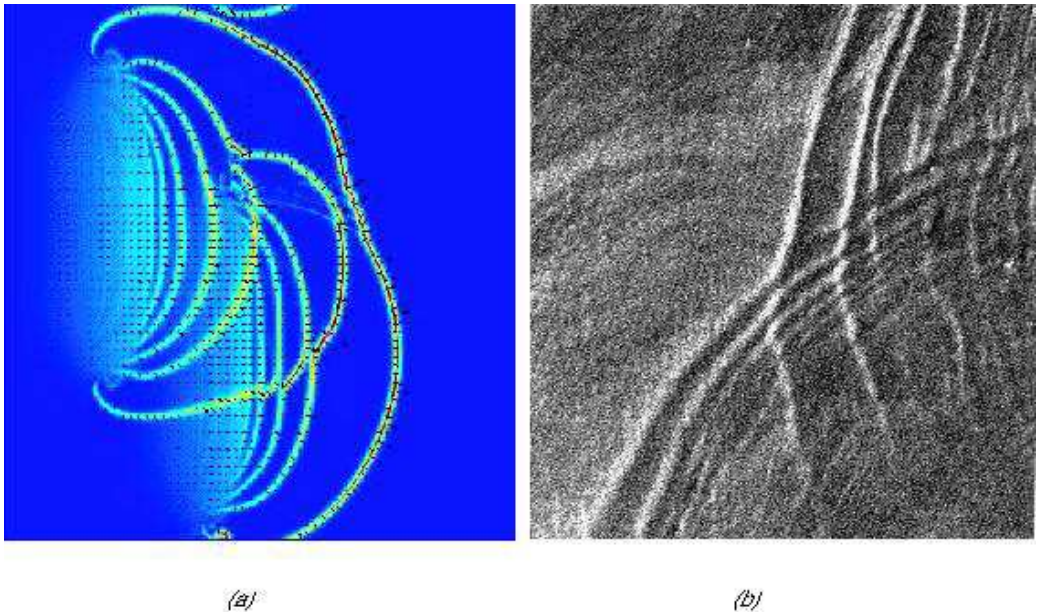}
\end{center}
\caption{\footnotesize Comparison of evolutionary EPDiff
solutions in two dimensions (a) and Synthetic Aperture Radar
observations by the Space Shuttle of internal waves in the South
China Sea (b). Both Figures show nonlinear reconnection occurring
in the wave train interaction as their characteristic feature. }
\label{strings_waves}
\end{figure}

Figure \ref{strings_waves} shows the striking reconnection phenomenon
seen in the nonlinear interaction between wave-trains, as simulated
by numerical solutions of the EPDiff equation and observed for
internal waves in the Ocean. Fig \ref{strings_waves}(a) shows a
frame taken from simulations of the initial value problem for
the EPDiff equation in two dimensions, excerpted from
\cite{HoSt2003}. (See also
\cite{HoPuSt2003}.) 

Fig \ref{strings_waves}(b) shows the interaction of two internal
wave trains propagating at the interface of different density
levels in the South China Sea, as observed from the Space
Shuttle using synthetic aperture radar, courtesy of A. Liu
(2002). Importantly, both Fig \ref{strings_waves}(a) and Fig
\ref{strings_waves}(b) show nonlinear reconnection occurring in
the wave train interaction as their characteristic feature.

Internal waves and these sorts of interactions are generally
thought to be described by the KP equation, and so any relations
among the KP equation, the EPDiff equation and the 2D CH
equation would be of great interest to explore; cf.
\cite{Liu1998}. The derivations of the KP equation and the CH
equation differ in the way the transverse motions are treated
in the asymptotics; so some difference in their solution behavior
is to be expected.

\section{Smoothness of the Lagrangian Equations}
\label{smoothness_section}

\paragraph{The One-Dimensional Case.} Based on a formal argument
given in \cite{HoMaRa1998a}, it was shown in \cite{Shkoller1998}
that when the CH equation (\ref{CHLP}) is transformed to
Lagrangian variables, it defines a smooth vector field (that is,
one obtains an evolution equation with no derivative loss). This
means that one can show using ODE methods that the initial value
problem is well-posed and one may also establish other
important properties of the equations when the data is
sufficiently smooth. 

As above, we write the relation between
$m$ and $u$ as $m = Q_{\rm op} u $, so that in the one dimensional
case, $Q_{\rm op}$ is the operator $Q_{\rm op} = {\rm Id} - \alpha ^2
\partial_{xx }$. We first recall how the equations
are transformed into Lagrangian variables. Introduce the one
parameter curve of diffeomorphisms $\eta (x_0,t)$ defined
implicitly by
%-----------------------------
\begin{equation} \label{defeta}
\frac{\partial}{\partial t} \eta (x_0,t) = u ( \eta (x_0,t), t ),
\end{equation}
%-----------------------------
so that $\eta$ is a geodesic in the group of diffeomorphisms of
$\mathbb{R}$ (or, with periodic boundary conditions, of the
circle $S ^1$) equipped with the right invariant metric equaling
the $H ^1$ metric at the identity.

We compute the second time derivative of $\eta$ in a straightforward
way by differentiating (\ref{defeta}) using the chain rule:
\[
\frac{\partial ^2 \eta }{\partial t^2 } =
uu_x + \frac{\partial u}{\partial t} 
\,.
\]
Acting on this equation with $Q_{\rm op}$ and using the definition 
$m = Q_{\rm op} u $ yields
\begin{align*}
Q_{\rm op} \frac{\partial ^2 \eta }{\partial t ^2 } 
& =
Q_{\rm op}(uu_x) - u\partial_x(Q_{\rm op}u)
+  u m_x
+ \frac{\partial m }{\partial t }
 \\
& = 
[Q_{\rm op}\,,\,u\partial_x]u - 2 m u_x
 \\
& = 
- 3 {\alpha}^2 u_x u_{xx}  -  2m u_x
 \\
& = 
- {\alpha}^2 u_x u_{xx} - 2 u u_x 
\,,
\end{align*}
where the third step uses the commutator relation calculated from the
product rule,
\[
[Q_{\rm op}\,,\,u\partial_x]u = - 3 {\alpha}^2 u_x u_{xx}
\,.
\]
Hence, the preceding equation becomes
\begin{equation} \label{chlag}
\frac{\partial ^2 \eta }{\partial t ^2 }
= -\,\frac{1}{2}\,Q_{\rm op} ^{-1}\, \partial_x
\left( \alpha^2 u_x^2 + 2u^2 \right).
\end{equation}

The important point about this equation is that the right hand side
has no derivative loss. That is, if $u$ is in the Sobolev space
$H^s$ for $s > 5/2$, then the right hand is also in the same space.
Regarding the right hand side as a function of $\eta$ and $\partial
\eta/\partial t $, we see that it is plausible that  the second order
evolution equation (\ref{chlag}) for $\eta$ defines a smooth ODE on
the group of $H ^s$ diffeomorphisms. (This argument requires the use
of, for example, weighted Sobolev spaces in the case $ x \in
\mathbb{R}$). 

The above is the essence of the argument given
in \cite{Shkoller1998}, Remark 3.5, which in turn makes use of the
type of arguments found in \cite{EbMa1970} for the incompressible
case and which shows, by a more careful argument, that the spray is
smooth if $s > 3/2$. However, one should note that the complete
argument is not quite so simple (just as in the case of incompressible
fluids). A subtilty arises because smoothness means as a function
of $\eta\,,\dot{\eta}$. Hence, one must express $u$ in terms of
$\eta$, namely through the relation
$u _t = \dot{\eta}_t \circ
\eta ^{-1}_t$, where the subscript $t$ here denotes that this
argument is held fixed, and is not a partial derivative. Doing this,
one sees that, while there is clearly no derivative loss, the right
hand side of (\ref{chlag}) does involve $\eta^{-1}_t$ and the map
$\eta_t \mapsto \eta ^{-1}_t$ is known to {\it not} be smooth (just as
in \cite{EbMa1970}). Nevertheless, the {\it combination} that appears
in (\ref{chlag}) is, quite remarkably, a smooth function of $\eta,
\dot{\eta}$.

\paragraph{The $n$-Dimensional Case.} The above argument readily
generalizes to the
$n$-dimensions, which we shall present in the case of $\mathbb{R}^n$
or the flat
$n$-torus
$\mathbb{T}^n$ for simplicity. Namely, we still have the
relation
\[
\frac{\partial}{\partial t} \eta (\mathbf{x}_0,t) 
= \mathbf{u} ( \eta (\mathbf{x}_0,t), t ),
\]
between $\eta$ and $\mathbf{u}$. Consequently, we may compute the
second partial time derivative of $\eta$ in the usual fashion
using the chain rule: 
\[
\frac{\partial ^2 \eta }{\partial t^2 } =
\mathbf{u} \cdot \nabla \mathbf{u} + \frac{\partial
\mathbf{u}}{\partial t}
\,. 
\]
Therefore, as in the one dimensional case, we get
\[
Q_{\rm op} \frac{\partial ^2 \eta }{\partial t ^2 }  =
\big[Q_{\rm op}\,,\,(\mathbf{u} \cdot \nabla)\big] \mathbf{u} 
+ \mathbf{u} \cdot \nabla \mathbf{m}
+ \frac{\partial \mathbf{m} }{\partial t }
\,.
\]
Calculating the commutator relation in $n$-dimensions gives
\[
\big[Q_{\rm op}\,,\,(\mathbf{u} \cdot \nabla)\big] \mathbf{u} 
= 
-\,
\alpha^2{\rm div}\Big( \nabla\mathbf{u}\cdot\nabla\mathbf{u}
+
\nabla\mathbf{u}\cdot\nabla\mathbf{u}^{\,T}  \Big)
+
\alpha^2(\nabla\mathbf{u})\cdot\nabla{\rm div}\,\mathbf{u}
\]
or, in components, 
\begin{equation}
\big(
\big[Q_{\rm op}\,,\,(\mathbf{u} \cdot \nabla)\big] \mathbf{u}
\big)\,_i =
-\,
\alpha^2\partial_k\Big(
u_{i,j}\,u_{j,k} + u_{i,j}\,u_{k,j} 
\Big)
+
\alpha^2(u_{i,j})\,\partial_j{\rm div}\,\mathbf{u}
\,,
\label{div-tensor-tau}
\end{equation}
with a sum on repeated indices.

Upon substituting the preceding commutator relation, the EPDiff
equation (\ref{EP-eqn-vec-intro}) and the vector calculus
identity 
%%%%%%%%%%%%%%%%%%%%%%%%%%%%%%%%%%%%%%%%%%%%%%%%%%%%%%%%%%%%%%%%%%%%%%%%
\begin{equation} \label{vect-id}
-\nabla\mathbf{u}^T\cdot\mathbf{m}
=
\alpha^2 \,{\rm div}\,\Big(
\nabla \mathbf{u}^{\,T}\cdot\nabla\mathbf{u}
\,\Big)
-
\nabla\Big(\frac{1}{2}|\mathbf{u}|^2  
+ \frac{\alpha^2}{2}|\nabla\mathbf{u}|^2  \,\Big)
\,,
\end{equation}
%%%%%%%%%%%%%%%%%%%%%%%%%%%%%%%%%%%%%%%%%%%%%%%%%%%%%%%%%%%%%%%%%%%%%%%%
then imply the $n-$dimensional result
\begin{align*}
Q_{\rm op} \frac{\partial ^2 \eta }{\partial t ^2 } 
& =
-\,
\alpha^2{\rm div}
\Big( \nabla\mathbf{u}\cdot\nabla\mathbf{u}
+
\nabla\mathbf{u}\cdot\nabla\mathbf{u}^{\,T}  
-
\nabla \mathbf{u}^{\,T}\cdot\nabla\mathbf{u}
- 
\nabla \mathbf{u}^{\,T}({\rm div}\,\mathbf{u})
+
\frac{1}{2}\,{\rm Id} |\nabla \mathbf{u}|^2
\Big)
\\  & \quad
- \mathbf{u}({\rm div}\,\mathbf{u})
- \frac{1}{2}\,\nabla|\mathbf{u}|^2
\,.
\end{align*}

This form of the EPDiff equation is useful for interpreting
some of its solution behavior. As in the one dimensional case, the
crucial point is that the right hand side involves at
most second derivatives of $\mathbf{u}$; so there is no
derivative loss in the overall expression. The precise
formulation of the smoothness result thus should hold, as in the
one-dimensional case, although, naturally the details are a bit
more complicated. We shall leave the technical details for
another publication.

A consequence is that the equations are well posed {\it for short
time} if the initial data is smooth enough and, hence, e.g., two
nearby smooth solutions can be joined by a unique geodesic.
Moreover, the solutions of the EPDiff equation are automatically
$C ^{\infty}$ in time. In the sections that follow, we will be
interested in nonsmooth data, which is in stark contrast to
the preceding discussion, which requires initial data that is at
least $C ^1$. 

\medskip
Remarkably, the same smoothness results hold for the case of the
LAE-$\alpha$ (Lagrangian averaged Euler) equations, a set of
incompressible equations in which small scale fluctuations are
averaged. One can view the LAE-$\alpha$ equations as the
incompressible version of the EPDiff equations. This smoothness
property for the LAE-$\alpha$ equations was shown by
\cite{Shkoller1998} for regions with no boundary and for regions
with boundary (for various boundary conditions), it was shown in 
\cite{MaRaSh2000}.  However, unlike the incompressible case, the
results apparently do not hold if $\alpha$ is zero (as also noted
in \cite{Shkoller1998}). This sort of smoothness result also
appears not to hold for many other equations, such as the KdV
equation, even though it too can be realized as Euler-Poincar\'e
equations on a Lie algebra, or as geodesics on a group, in this
case the Bott-Virasoro group, as explained in \cite{MaRa1999} and
references therein.

\paragraph{The Development of Singularities.} The smoothness
property just discussed does not preclude the development in
finite time of singular solutions from smooth localized initial
data as was indicated in Figure \ref{peakon_figure}. To capture
the local singularities in the EPDiff solution (either
verticality in slope, or discontinuities in its spatial
derivative) that develop in finite time from arbitrarily smooth
initial conditions, one must enlarge the solution class of
interest, by considering weak solutions. 

There are a number of papers on weak solutions of the CH equation
such as \cite{XiZh2000} that we will not survey here. We just
mention that the theory is not yet complete, as it is
still unknown in what sense one may define {\it global unique}
weak solutions to the CH equations in $H^1$. As discussed in
\cite{AlCaFeHoMa2001} for the CH equation, one most likely must
consider weak solutions in the {\it spacetime sense}.

The steepening lemma of \cite{CaHo1993} proves that in one dimension
any initial velocity distribution whose spatial profile has an
inflection point with negative slope (for example, any
antisymmetric smooth initial distribution of velocity on the real
line) will develop a vertical slope in finite time. Note that the
peakon solution  (\ref{u-peakon-ansatz-intro}) has no inflection
points, so it is not subject to the steepening lemma. However,
the steepening lemma underlies the mechanism for forming these
singular solutions, which are continuous but have discontinuous
spatial derivatives; they also lie in $H ^1$ and have finite
energy. We conclude that solutions with initial conditions in $H
^s$ with $s > (n/2) + 1$ go to infinity in the $H^s$ norm in
finite time, but remain in $H^1$ and presumably continue to
exist in a weak spacetime sense for all time in $H^1$.

Numerical evidence in higher dimensions and the inverse
scattering solution for the CH equation in one dimension (the
latter has {\it only} discrete eigenvalues, corresponding to
peakons) both show that the singular solutions completely dominate
the time-asymptotic dynamics of the initial value problem (IVP).
This singular IVP behavior is one of the main discoveries of
\cite{CaHo1993}. This singular behavior has drawn a great deal
of mathematical interest to the CH equation and its relatives,
such as EPDiff. The other properties of CH --- its complete
integrability, inverse scattering transform, connections to
algebraic geometry and elliptical billiards, bi-Hamiltonian
structure, etc. --- are all interesting, too. However, the
requirement of dealing with singularity as its main solution
phenomenon is the primary aspect of CH (and EPDiff). We aim to
show that many of the properties of these singular solutions of
CH and EPDiff are  captured by recognizing that the singular
solution ansatz itself is a momentum map. This momentum map
property explains, for example, why the singular solutions 
(\ref{m-ansatz-intro}) form an invariant manifold for any value
of $N$. 
\medskip

In one dimension, the complete integrability of the CH equation as
a Hamiltonian system and its soliton paradigm completely explain
the emergence of peakons in the CH dynamics. Namely, their
emergence reveals the initial condition's soliton (peakon)
content. However, beyond one dimension, we do not have an explicit
mechanism for explaining why {\it only} singular solution
behavior emerges in numerical simulations. One hopes  that
eventually a theory will be developed for explaining this
singular solution phenomenon in  higher dimensions. Such a theory
might, for example, parallel the well-known explanation of the
formation of shocks for hyperbolic partial differential
equations. (Note, however, that EPDiff is not hyperbolic, because
the relation $\mathbf{u}=G*\mathbf{m}$ between its velocity and
momentum is nonlocal.) 
\medskip

In the remainder of this work, we shall focus our attention on the
momentum map properties of the invariant manifold of singular
solutions (\ref{m-ansatz-intro}) of the EPDiff equation.

\section{The Singular Solution Momentum Map}
\label{SSMomMap_section}

\paragraph{The Momentum Ansatz \textup{(\ref{m-ansatz-intro})} is
a Momentum Map.} The purpose of this section is to show that the
solution ansatz \textup{(\ref{m-ansatz-intro})} for the momentum
vector in the EPDiff equation (\ref{EP-eqn-vec-intro}) defines a
momentum map for the action of the group of diffeomorphisms on
the support sets of the Dirac delta functions. These support
sets are the analogs of points on the real line for the CH
equation in one dimension. They are points, curves, or surfaces
in $\mathbb{R}^n$ for the EPDiff equation in $n-$dimensions.

This result, as we shall discuss in greater detail later, shows that
the solution ansatz \textup{(\ref{m-ansatz-intro})} fits
naturally into the scheme of Clebsch, or canonical variables in
the sense advocated by \cite{MaWe1983} as well as showing
that these singular solutions evolve on special coadjoint orbits
for the diffeomorphism group. 

One can summarize by saying
that the map that implements the canonical
$(\mathbf{Q},
\mathbf{P})$ variables in terms of singular solutions is a
(cotangent bundle) momentum  map. Such momentum
maps are Poisson maps; so the canonical Hamiltonian
nature of the dynamical equations for
$(\mathbf{Q}, \mathbf{P})$ fits into a general theory which also
provides a framework for suggesting other avenues of
investigation.

\begin{theorem}\label{mom-map}
The momentum ansatz \textup{(\ref{m-ansatz-intro})} for
measure-valued solutions of the \textup{EPDiff} equation
\textup{(\ref{EP-eqn-vec-intro})}, defines an equivariant
momentum map
\[
\mathbf{J}_{\rm Sing}: T ^{\ast} \operatorname{Emb}(S, \mathbb{R}^n)
\rightarrow
\mathfrak{X}(\mathbb{R}^n)^{\ast}
\]
that we will call the {\bfi singular solution momentum map}.
\end{theorem}

We shall explain the notation used in this statement in the
course of the proof. Right away, however, we note that the sense
of ``defines'' is that expressing $\mathbf{m}$ in terms of
$\mathbf{Q}, \mathbf{P}$ (which are, in turn, functions of $s$)
can be regarded as a map from the space of $(\mathbf{Q}(s),
\mathbf{P} (s))$ to the space of $\mathbf{m}$'s. This will turn
out to be the Lagrange-to-Euler map for the fluid description of
the singular solutions. 
\medskip

 We shall give two proofs of this result from two
rather different points of view. The first proof below uses the
formula for a momentum map for a cotangent lifted action, while
the second proof focuses on a Poisson bracket computation. Each
proof also explains the context in which one has a momentum map.
(See \cite{MaRa1999} for general background on momentum maps.)
\medskip

\noindent{\bf First Proof.} For simplicity and without loss of
generality, let us take $N = 1$ and so suppress the index $a$.
That is, we shall take the case of an isolated singular
solution. As the proof will show, this is not a real
restriction. 
\medskip

To set the notation, fix a
$k$-dimensional manifold
$S$ with a given volume element and whose points are denoted $s
\in S$. Let $\operatorname{Emb}(S, \mathbb{R}^n)$ denote the
set of smooth embeddings $\mathbf{Q}: S \rightarrow \mathbb{R}^n$.
(If the EPDiff equations are taken on a manifold $M$, replace
$\mathbb{R}^n $ with $M$.) Under appropriate technical
conditions, which we shall just treat formally here,
$\operatorname{Emb}(S, \mathbb{R}^n)$ is a smooth manifold. (See,
for example, \cite{EbMa1970} and \cite{MaHu1983} for a discussion
and references.)
\medskip

The tangent space $T _{\mathbf{Q}} \operatorname{Emb}(S,
\mathbb{R}^n)$ to $\operatorname{Emb}(S, \mathbb{R}^n)$ at the
point $\mathbf{Q} \in \operatorname{Emb}(S, \mathbb{R}^n)$ is
given by the space of {\bfi material velocity fields}, namely the
linear space of maps
$\mathbf{V}: S \rightarrow \mathbb{R}^n$ that are vector fields
over the map $\mathbf{Q}$.  The dual space to this space will be
identified with the space of one-form densities over
$\mathbf{Q}$, which we shall regard as maps $\mathbf{P}: S
\rightarrow \left(\mathbb{R}^n\right) ^{\ast}$. In summary, the
cotangent bundle $T ^{\ast} \operatorname{Emb}(S, \mathbb{R}^n)$
is identified with the space of pairs of maps $\left( \mathbf{Q},
\mathbf{P} \right)$.
\medskip

These give us the domain space for the singular solution momentum
map. Now we consider the action of the symmetry group.  Consider
the group $\mathfrak{G} = \operatorname{Diff}$ of diffeomorphisms
of the space $\mathfrak{S}$ in which the EPDiff equations are
operating, concretely in our case $\mathbb{R}^n$. Let it act on
$\mathfrak{S}$ by composition on the {\it left}. Namely for $\eta
\in \operatorname{Diff} (\mathbb{R}^n)$, we let 
%-----------------------------
\begin{equation} \label{action}
\eta \cdot \mathbf{Q} = \eta \circ \mathbf{Q}.
\end{equation} 
%-----------------------------
Now lift this action to the cotangent bundle $T ^{\ast} \operatorname{Emb}(S, \mathbb{R}^n)$
in the standard way (see, for instance, \cite{MaRa1999} for this
construction). This lifted action is a symplectic (and hence Poisson)
action and has an equivariant momentum map. {\it We claim that
this momentum map is precisely given by the ansatz
\textup{(\ref{m-ansatz-intro})}.}

To see this, we just need to recall and then apply the general formula
for the momentum map associated with an action of a general Lie group
$\mathfrak{G}$ on a configuration manifold $Q$ and cotangent lifted to
$T^{\ast}Q$.

First let us recall the general formula. Namely, the momentum map  is
the map $\mathbf{J}: T^{\ast}Q \rightarrow
\mathfrak{g}^\ast$  ($\mathfrak{g}^\ast$ denotes the dual
of the Lie algebra $\mathfrak{g}$ of $\mathfrak{G}$) defined by
\begin{equation} \label{momentummap}
\mathbf{J} (\alpha _q) \cdot \xi = \left\langle \alpha _q, \xi_Q (q)
\right\rangle,
\end{equation}
where $\alpha_q \in T ^{\ast} _q Q $ and $\xi \in \mathfrak{g}$, 
where $\xi _Q $ is the infinitesimal generator of the action of
$\mathfrak{G}$ on $Q$ associated to the Lie algebra element $\xi$,
and where
$\left\langle \alpha _q, \xi_Q (q)
\right\rangle$ is the natural pairing of an element of $T ^{\ast}_q
Q $ with an element of $T _q Q $.  

Now we apply this formula to the special case in which 
the group $\mathfrak{G}$ is the diffeomorphism group
$\operatorname{Diff} (\mathbb{R}^n)$, the manifold $Q$ is
$\operatorname{Emb}(S, \mathbb{R}^n)$ and where the action of
the group on
$\operatorname{Emb}(S, \mathbb{R}^n)$ is given by
(\ref{action}). The sense in which the Lie algebra of
$\mathfrak{G} = \operatorname{Diff}$ is the space
$\mathfrak{g} = \mathfrak{X}$ of vector fields is
well-understood. Hence, its dual is naturally regarded as
the space of one-form densities. The momentum map is thus a map
$\mathbf{J}: T ^{\ast}
\operatorname{Emb}(S, \mathbb{R}^n)
\rightarrow \mathfrak{X}^{\ast}$.
\medskip

With $\mathbf{J}$ given by (\ref{momentummap}), we just need to
work out this formula. First, we shall work out the infinitesimal
generators. Let $X \in \mathfrak{X}$ be a Lie algebra element. By
differentiating the action (\ref{action}) with respect to $\eta$
in the direction of $X$ at the identity element we find that the
infinitesimal generator is given by
\[
X _{\operatorname{Emb}(S, \mathbb{R}^n)} (\mathbf{Q}) 
= X \circ \mathbf{Q}.
\]
Thus, taking $\alpha _q$ to be the cotangent vector
$(\mathbf{Q}, \mathbf{P})$, equation
(\ref{momentummap}) gives
\begin{align*}
\left\langle \mathbf{J} (\mathbf{Q}, \mathbf{P} ),
X \right\rangle & = \left\langle (\mathbf{Q}, \mathbf{P}),
X \circ \mathbf{Q} \right\rangle \\
& = \int_{S} P _i(s) X ^i (\mathbf{Q}(s)) d^k s .
\end{align*}
On the other hand, note that the right hand side of
(\ref{m-ansatz-intro}) (again with the index $a$ suppressed,
and with $t$ suppressed as well), when paired with the
Lie algebra element $X$ is
\begin{align*}
\left\langle \int _S \mathbf{P}(s)\,
\delta \left( \mathbf{x}-\mathbf{Q}(s) \right) d^k s,
X \right\rangle & = \int _{\mathbb{R}^n} 
\int _S  \left( P _i (s)\,
\delta \left( \mathbf{x}-\mathbf{Q}(s) \right) d ^k s \right)
X ^i (\mathbf{x}) d ^n x \\
& = \int _S P _i (s) X ^i (\mathbf{Q} (s) d ^k s.
\end{align*}
This shows that the expression given by
(\ref{m-ansatz-intro}) is equal to $\mathbf{J}$ and 
so the result is proved. \quad $\blacksquare$

\medskip
\noindent{\bf Second Proof.} As is standard (see, for example,
\cite{MaRa1999}), one can characterize momentum maps
by means of the following relation, required to hold for all functions
$F$ on $ T^{\ast} \operatorname{Emb}(S, \mathbb{R}^n)$; that is,
functions of $\mathbf{Q}$ and $\mathbf{P}$:
\begin{equation} \label{mom_map_Poisson}
\left\{ F, \left\langle \mathbf{J}, \xi \right\rangle \right\}
=  \xi_P [F]
\,.
\end{equation}
In  our case, we shall take $\mathbf{J}$ to be given by the
solution ansatz and verify that it satisfies this relation.
To do so, let $\xi\in\mathfrak{X}$ so that the left side
of (\ref{mom_map_Poisson}) becomes
\[
\left\{ F,
\int _S P _i (s) \xi\,^i (\mathbf{Q} (s)) d\, ^k s
\right\}
=
\int _S \left[
\frac{\delta F}{\delta Q^i }\xi\,^i (\mathbf{Q} (s))
-
P _i (s)\frac{\delta F}{\delta P_j }
\frac{\delta}{\delta Q^j }
\xi\,^i (\mathbf{Q} (s))
\right]d\, ^k s
\,.
\]
On the other hand, one can directly compute from the definitions
that the infinitesimal generator of the action on the
space
$T ^{\ast}  \operatorname{Emb}(S, \mathbb{R}^n)$ corresponding to the
vector field $\xi^i(\mathbf{x})\frac{\partial }{\partial Q ^i}$ (a
Lie algebra element), is given by (see \cite{MaRa1999}, formula
(12.1.14)):
\[
\delta\mathbf{Q}
= \xi \circ \mathbf{Q}
\,,\quad
\delta\mathbf{P}
=  -\,P _i (s)\frac{\partial}{\partial \mathbf{Q} }
\xi\,^i(\mathbf{Q} (s)),
\]
which verifies that (\ref{mom_map_Poisson}) holds.
\medskip

An important {\color{blue}element} left out in this proof so far is that it
does not make clear that the momentum map is {\it equivariant}, a condition
needed for the momentum map to be Poisson. The first proof took
care of this automatically since momentum maps for cotangent
lifted  actions are always equivariant and hence Poisson.
\medskip

Thus, to complete the second proof, we need to check  directly
that the momentum map is equivariant. Actually, we shall only
check that it is infinitesimally invariant by showing that it is
a Poisson map from $T ^{\ast} \operatorname{Emb}(S,
\mathbb{R}^n)$ to the space of $\mathbf{m}$'s (the dual of the
Lie algebra of
$\mathfrak{X}$) with its Lie-Poisson bracket. This sort of
approach to characterize equivariant momentum maps is discussed
in an interesting way in \cite{Weinstein2002}.
\medskip

The following computation accomplishes this methodology by
showing that the singular solution momentum map is Poisson.
 
Indeed, we use the canonical Poisson brackets
for $\{\mathbf{P}\},\,\{\mathbf{Q}\}$ and apply the chain rule to
compute
$\big\{m_i(\mathbf{x}),m_j(\mathbf{y})\big\}$, with notation
$\delta\,^\prime_k(\mathbf{y})
\equiv\partial\delta(\mathbf{y})/\partial{y^k}$. We get

\begin{eqnarray}\label{m-m-bracket1}
\big\{m_i(\mathbf{x}),m_j(\mathbf{y})\big\}\hspace{-1in}&&
\nonumber\\
&=&
\bigg\{\sum_{a=1}^N\!\int\!\!ds\, P_i^a(s,t)
\, \delta (\mathbf{x}-\mathbf{Q}^a(s,t))
\,,\,
\sum_{b=1}^N \!\int\!\!ds^\prime P_j^b(s^\prime,t)\,
\delta (\mathbf{y}-\mathbf{Q}^b(s^\prime,t))\bigg\}
\nonumber\\
&=&
\sum_{a,b=1}^N\int\!\!\!\int\!\!dsds^\prime
\bigg[ \{ P_i^a(s),P_j^b(s^\prime) \}
\,\delta (\mathbf{x}-\mathbf{Q}^a(s))
\,\delta (\mathbf{y}-\mathbf{Q}^b(s^\prime))
\nonumber\\
&&-\,\{ P_i^a(s),Q_k^b(s^\prime) \} P_j^b(s^\prime)
\,\delta (\mathbf{x}-\mathbf{Q}^a(s))
\,\delta\,^\prime_k(\mathbf{y}-\mathbf{Q}^b(s^\prime))
\nonumber\\
&&-\,\{ Q_k^a(s),P_j^b(s^\prime) \} P_i^a(s)
\,\delta\,^\prime_k (\mathbf{x}-\mathbf{Q}^a(s))
\,\delta (\mathbf{y}-\mathbf{Q}^b(s^\prime))
\nonumber\\
&&+\,\{ Q_k^a(s),Q_\ell^b(s^\prime) \} P_i^a(s)P_j^b(s^\prime)
\,\delta\,^\prime_k (\mathbf{x}-\mathbf{Q}^a(s))
\,\delta\,^\prime_\ell (\mathbf{y}-\mathbf{Q}^b(s^\prime))
\bigg]
\,.
\nonumber
\end{eqnarray}
Substituting the canonical Poisson bracket relations
\begin{align*}
\{ P_i^a(s),P_j^b(s^\prime) \} & = 0 \\
\{ Q_k^a(s),Q_\ell^b(s^\prime) \} & = 0, \quad \mbox{and} \; \\
\{ Q_k^a(s),P_j^b(s^\prime) \} & =
\delta^{ab}\delta_{kj}\delta(s-s^\prime)
\end{align*}
into the preceding computation yields,
\begin{eqnarray}\label{m-m-bracket2}
\big\{m_i(\mathbf{x}),m_j(\mathbf{y})\big\}\hspace{-1in}&&
\nonumber\\
&=&
\bigg\{\sum_{a=1}^N\!\int\!\!ds P_i^a(s,t)
\, \delta (\mathbf{x}-\mathbf{Q}^a(s,t))
\,,\,
\sum_{b=1}^N \!\int\!\!ds^\prime P_j^b(s^\prime,t)\,
\delta (\mathbf{y}-\mathbf{Q}^b(s^\prime,t))\bigg\}
\nonumber\\
&=&
\sum_{a=1}^N\int\!\!ds P_j^a(s)
\,\delta (\mathbf{x}-\mathbf{Q}^a(s))
\,\delta\,^\prime_i (\mathbf{y}-\mathbf{Q}^a(s))
\nonumber\\
&&-\sum_{a=1}^N\int\!\!ds P_i^a(s)
\,\delta\,^\prime_j (\mathbf{x}-\mathbf{Q}^a(s))
\,\delta (\mathbf{y}-\mathbf{Q}^a(s))
\nonumber\\
&=&
-\,
\Big(
m_j(\mathbf{x})\frac{\partial}{\partial x^i}
+
\frac{\partial}{\partial x\,^j}\,m_i(\mathbf{x})\Big)
\delta(\mathbf{x}-\mathbf{y})
\,.\nonumber
\end{eqnarray}

\noindent
Thus,
\begin{eqnarray}\label{momentum-map-bracket1}
\big\{m_i(\mathbf{x})\,,\,m_j(\mathbf{y})\big\}
=
-\,
\Big(
m_j(\mathbf{x})\frac{\partial}{\partial x^i}
+
\frac{\partial}{\partial x\,^j}\,m_i(\mathbf{x})\Big)
\delta(\mathbf{x}-\mathbf{y})
\,,\label{LP-bracket}
\end{eqnarray}
which is readily checked to be the Lie-Poisson bracket on the space
of $m$'s. This completes the second proof of theorem. \quad
$\blacksquare$
\medskip

Each of these proofs has shown the following basic fact.

\begin{corollary} \label{Poisson_mom-map}
The singular solution momentum map defined by the singular
solution ansatz, namely,
\[
\mathbf{J}_{\rm Sing}: 
T ^{\ast} \operatorname{Emb}(S, \mathbb{R}^n)
\rightarrow
\mathfrak{X}(\mathbb{R}^n)^{\ast}
\]
is a Poisson map from the canonical Poisson structure on $T
^{\ast} \operatorname{Emb}(S, \mathbb{R}^n)$ to the Lie-Poisson
structure on $\mathfrak{X}(\mathbb{R}^n)^{\ast}$.
\end{corollary}

This is perhaps the most basic property of the singular solution
momentum map. Some of its more sophisticated properties are
outlined in the following section.

\paragraph{Pulling Back the Equations.}
Since the solution ansatz (\ref{m-ansatz-intro}) has been shown
in the preceding Corollary to be a Poisson map, the pull back of
the Hamiltonian from $\mathfrak{X}^{\ast}$ to $T ^{\ast}
\operatorname{Emb}(S, \mathbb{R}^n)$ gives equations of motion on
the latter space that project to the equations on $\mathfrak{X}
^{\ast}$. This is why the functions
$\mathbf{Q}^a(s,t)$ and $\mathbf{P}^a(s,t)$ 
satisfy canonical Hamiltonian equations. Note that the coordinate
$s\in{\mathbb{R}}^{k}$ that labels these functions is a Lagrangian
coordinate. 

In terms of the pairing
\begin{equation}
\langle\cdot\,,\,\cdot\rangle:
\,\mathfrak{g}^*\times\mathfrak{g}\to{\mathbb{R}}
\,,
\end{equation}
between the Lie algebra $\mathfrak{g}$ (vector fields in $\mathbb{R}^n$)
and its dual $\mathfrak{g}^*$  (one-form densities in $\mathbb{R}^n$), the
following relation holds for measure-valued solutions
under the momentum map (\ref{m-ansatz-intro}),
\begin{eqnarray}\label{momentum-map-relation}
\langle\mathbf{m}\,,\,\mathbf{u}\rangle
&=&
\int \mathbf{m}\,\cdot\,\mathbf{u}\,d\,^n\mathbf{x}
\,,\quad\hbox{$L^2$ pairing for }
\mathbf{m}\,\&\,\mathbf{u}\in{\mathbb{R}^n},
\nonumber\\
&=&
\!\int\!\!\!\!\int\!\!\sum_{a\,,\,b=1}^{N}
\big(\mathbf{P}^a(s,t)\cdot\mathbf{P}^b(s^{\prime},t)\big)
\,G\big(\mathbf{Q}^a(s,t)-\mathbf{Q}^{\,b}(s^{\prime},t)\big)
\,ds\,ds^{\prime}
\nonumber\\
&=&
\!\int\!\!\sum_{a=1}^{N}
\mathbf{P}^a(s,t)\cdot\frac{\partial\mathbf{Q}^a(s,t)}{\partial t}
\,ds
\nonumber\\
&\equiv&
\langle\!\langle\mathbf{P}\,,\,\mathbf{\dot{Q}}\rangle\!\rangle,
\end{eqnarray}
which is the natural pairing between the points $(\mathbf{Q},
\mathbf{P})
\in  T^{\ast} \operatorname{Emb}(S, \mathbb{R}^n)$ and $(\mathbf{Q},
\dot{\mathbf{Q}}) \in T \operatorname{Emb}(S, \mathbb{R}^n)$.  

The pull-back of the Hamiltonian  $H[\mathbf{m}]$ defined on the dual
of the Lie algebra $\mathfrak{g}^*$, to $ T^{\ast}
\operatorname{Emb}(S, \mathbb{R}^n)$ is easily seen to be consistent
with what we had before:
\begin{equation}
H[\mathbf{m}]
\equiv
\frac{1}{2}\langle\mathbf{m}\,,\,G*\mathbf{m}\rangle
=
\frac{1}{2}\langle\!\langle\mathbf{P}\,,\,G*\mathbf{P}\rangle\!\rangle
\equiv
H_N[\mathbf{P},\mathbf{Q}]
\,.
\label{geodesic-ham}
\end{equation}

In summary, in concert with the Poisson nature of the singular
solution momentum map, we see that the singular solutions in
terms of $\mathbf{Q}$ and $\mathbf{P}$ satisfy Hamiltonian
equations and also define an invariant solution set for the
EPDiff equations. In fact, this invariant solution set is a
special coadjoint orbit for the diffeomorphism group, as we
shall discuss in the next section.

\paragraph{Smoothness.} It would be extremely interesting if the
smoothness properties explored in \S\ref{smoothness_section} were
also valid on the space $T^{\ast}
\operatorname{Emb}(S, \mathbb{R}^n)$. We hope to explore this
point in future publications.

\section{The Geometry of the Momentum Map} \label{geom_mommap}

In this section we explore the geometry of the singular
solution momentum map discussed in \S\ref{SSMomMap_section} in
a little more detail. The main idea may be stated as follows: {\it
simply apply all of the ideas given in \textup{\cite{MaWe1983}}
in a systematic way to the current setting.}

\paragraph{Coadjoint Orbits.} The first claim is that {\it the
image of the singular solution momentum map is a coadjoint orbit
in $\mathfrak{X} ^{\ast}$.}  This means that (perhaps modulo some
issues of connectedness and smoothness, which we do not consider
here) the solution anzatz given by  (\ref{m-ansatz-intro})
defines a coadjoint orbit in the space of all one-form densities,
regarded as the dual of the Lie algebra of the diffeomorphism
group. 

These coadjoint orbits should be thought of as singular
orbits---that is, due to their special nature, they are not
generic. Also, this explains why the singular solutions
form dynamically invariant sets---it is because they are
coadjoint orbits, which are {\it symplectic submanifolds} of the
Lie-Poisson manifold $\mathfrak{X}(\mathbb{R}^n)^{\ast}$.

 The idea of the proof of this is simply this: whenever one has an
equivariant momentum map $\mathbf{J}: P \rightarrow
\mathfrak{g}^\ast$ for the action of a group $G$ on a symplectic
or Poisson manifold $P$, and that action is transitive, then the
image of $\mathbf{J}$ is an orbit (or at least a piece of an
orbit). This general result, due to Kostant, is stated more
precisely in
\cite{MaRa1999}, Theorem 14.4.5. Roughly speaking, the reason
that transitivity holds in our case is because one can ``move
the manifolds $S$ around at will'' using diffeomorphisms.

\paragraph{Symplectic Structure on Orbits.} Recall (from, for
example, \cite{MaRa1999}), the general formula for the symplectic
structure on coadjoint orbits:
\begin{equation} \label{coadjoint_sympl}
\Omega _\mu \left( \xi _{\mathfrak{g}^\ast} (\mu), \eta
_{\mathfrak{g}^\ast } (\mu) \right)
=  \left\langle \mu,  [ \xi, \eta ] \right\rangle,
\end{equation}
where $\mu \in \mathfrak{g}^\ast$ is a chosen point on an
orbit and where $\xi, \eta$ are elements of $\mathfrak{g}$.
We use a plus sign in this formula since we are dealing with
orbits for the {\it right action}.
\medskip

Just as in
\cite{MaWe1983}, this leads to an explicit formula for the
coadjoint orbit symplectic structure in the case of 
$\operatorname{Diff}$. In the present case, it is a particularly
simple and transparent formula.

Recall that in the case of incompressible fluid mechanics, this
procedure leads naturally to the symplectic (and Poisson) structure
for many interesting singular coadjoint orbits, such as point vortices
in the plane, vortex patches, vortex blobs (closely related to the
planar LAE-$\alpha$ equations) and vortex filaments. 
 \medskip

For the case of the diffeomorphism group, let
$\mathcal{O}_{\mathbf{m}}$ denote the coadjoint orbit through the
point $\mathbf{m} \in \mathfrak{X}^{\ast} (\mathbb{R}^n)$.

\begin{theorem} The symplectic structure   $\Omega
_{\mathbf{m}}$ on $ T_{\mathbf{m}} \mathcal{O}_{\mathbf{m}} $ is
given by
\[
\Omega_{\mathbf{m}} (\pounds_{u_1} \mathbf{m}, \pounds_{u_{2}}
\mathbf{m} ) =  - \int  \left\langle \mathbf{m},  [u_1, u_2 ]
\right\rangle \, d^n x.
\]
\end{theorem}

\begin{proof}
We simply plug into the general Kirillov-Kostant-Souriau
formula  (\ref{coadjoint_sympl}) for the symplectic structure on
coadjoint orbits. (As noted above, there is a $ + $ sign, since
we are dealing with a {\it right\/} invariant system). The only
thing needing explanation is that our Lie algebra convention 
always uses the left Lie bracket. For
$\operatorname{Diff}$, this is the {\it negative\/} of
the usual Lie bracket as is explained in \cite{MaRa1999}.
\end{proof}

\paragraph{Dual Pairs.} The singular solution momentum map
 $\mathbf{J}_{\rm Sing}: 
T^{\ast} \operatorname{Emb}(S,\mathbb{R}^n)
\rightarrow
\mathfrak{X} (\mathbb{R}^n)^{\ast}$ 
forms one leg of a dual pair. The point is that
there is another group that acts on $
\operatorname{Emb}(S,
\mathbb{R}^n)$, namely the group 
$\operatorname{Diff} (S)$ of diffeomorphisms of
$S$, which acts on the {\it right}, while
$\operatorname{Diff}(\mathbb{R}^n)$ acted on it by composition on
the {\it left} (and which gave rise to our singular solution
momentum map). This momentum map will be denoted
$\mathbf{J}_S: T ^{\ast} 
\operatorname{Emb}(S, \mathbb{R}^n) \rightarrow
\mathfrak{X}(S)^{\ast}$. The dual pair arises for reasons that
are very similar to those in \cite{MaWe1983}. See the following
diagram.
\bigskip 

\begin{picture}(150,100)(-70,0)%  
\put(100,75){$T^{\ast} \operatorname{Emb}(S,\mathbb{R}^n)$} 
%top label

\put(78,50){$\mathbf{J}_{\rm Sing}$}        
%left label

\put(160,50){$\mathbf{J}_S$}   
%right arrow label

\put(72,15){$\mathfrak{X} (\mathbb{R}^n)^{\ast}$}       
%left bottom label

\put(170,15){$\mathfrak{X}(S)^{\ast}$}       
%right bottom label

\put(130,70){\vector(-1, -1){40}}  
% left slanted arrow

\put(135,70){\vector(1,-1){40}}  
% right slanted arrow

\end{picture}

We also note that this framework shows that the parameterization
of the singular solutions in terms of $\mathbf{Q}$ and
$\mathbf{P}$ are exactly Clebsch variables in the sense given in
\cite{MaWe1983}. Also notice that when we write the singular
solutions in $\mathbf{Q}$-$\mathbf{P}$ space, we are
finding solutions that are {\it collective} and so all the
properties of collectivization are valid. See \cite{MaRa1999}
for a general discussion and references to the original work of
Guillemin and Sternberg on this topic.

\section{Challenges, Future Directions and Speculations}
\label{future_section}

\paragraph{Numerical Issues: Geometric Integrators.} The computations of
Martin Staley that were illustrated in this paper and that are discussed in
\cite{HoSt2003}, make use of both mimetic differencing and
reversibility in a critical way and this is important for good
numerical resolution. In other words, integrators that respect the basic
geometry underlying the problem seem to play an key role. It would be
interesting to pursue this aspect further and also incorporate discrete
exterior calculus and variational multisymplectic integration methods (see 
\cite{DeHiLeMa2003} as well as \cite{MaPaSh1998} and \cite{LeMaOrWe2003}).

\paragraph{Analytical Issues: Geodesic Incompleteness of $H^1$ EPDiff.} The
emergence in finite time of singular solutions from smooth
initial data observed numerically in \cite{HoSt2003} indicates
that the diffeomorphism group with respect to the right invariant $H^1$
metric is {\it geodesically incomplete} when the diffeomorphism
group has the $H ^s $ topology, $s> (n/2) + 1$. The degree of its geodesic
incompleteness is not known, but we suspect that almost all EPDiff
geodesics in $H^1$ cannot be extended indefinitely. This certainly holds in
one spatial dimension, where the discreteness of the CH isospectrum implies
that asymptotically in time the CH solution arising from any confined
initial velocity data consists {\em only} of peakons. {\it It is an
important challange to find a context in which one can put the $H ^1$
topology on the diffeomorphism group and reestablish geodesic
completeness}. The numerics suggests that this might be possible, while
known existence theorems, even for the CH equation are not yet capable of
showing this---to the best of our knowledge.

\paragraph{Reversible Reconnections of the Singular EPDiff
Solutions.} EPDiff is a reversible equation,  and the collisions
of its peakon solutions on the line $\mathbb{R}^1$ (or the
circle $S^1$) are known to be reversible. In principle, the
reconnections of the singular EPDiff solutions observed
numerically in \cite{HoSt2003} in periodic domains
$\mathbb{T}^2$ and $\mathbb{T}^3$ must also be reversible.
Reversibility of its reconnections distinguishes the singular
solutions of EPDiff from vortex fluid solutions and shocks in
fluids, whose reconnections apparently require dissipation and
so, are not reversible. The mimetic finite differencing scheme
used for the numerical computation of EPDiff solutions in
\cite{HoSt2003} was indeed found to be reversible for overtaking
collisions, but it was found to be only approximately reversible
for head-on collisions, which are much more challenging for
numerical integration schemes. 

\paragraph{Applications of EPDiff Singular Solutions in Image
Processing.} The singular EPDiff solutions correspond to
outlines (or cartoons) of images in applications of geodesic
flow for the template, or pattern matching approach. The
dynamics of the singular EPDiff solutions described by the
momentum map \textup{(\ref{m-ansatz-intro})} introduces the
paradigm of soliton collisions into the mechanics and analysis
of image processing by template matching. (See
\cite{HoTrYo2003} for more discussions of this new paradigm for
image processing.) The reversibility of the collisions among
singular solutions and their reconnections under EPDiff flow
assures the preservation of the information contained in the
image outlines. In addition, the invariance of the manifold of
$N$ singular solutions under EPDiff assures that the fidelity of
the image is preserved in the sense of approximation theory.
That is, an $N$ soliton approximation of the image outlines
remains so, throughout the EPDiff flow. A natural approach for
numerically simulating EPDiff flows in image processing is to use
multisymplectic algorithms. The preservation of the space-time
multisymplectic form by these algorithms introduces an
initial-value, final-value formulation of the numerical solution
procedure that is natural for template matching. 

\paragraph{Rigorous Poisson Structures.} In \cite{VaMa2003}, the
question of the (rigorous) Poisson nature of the time $t$ map of
the flow of the Euler equations for an ideal fluid in appropriate
Sobolev spaces is explored. Given the smoothness
properties in \S\ref{smoothness_section}, it seems
reasonable that similar properties should also hold for the
EPDiff equations. However, as mentioned earlier, these
smoothness properties do not preclude the emergence of singular
solutions from smooth initial data in finite time, because of
the possibility for geodesic incompleteness. 

\paragraph{Other Groups.} The general setting of this paper
suggests that perhaps one should look for similar measure valued
or singular solutions associated with other problems, including
geodesic flows on the group of symplectic diffeomorphisms
(relevant for plasma physics, as in \cite{MaWe1982}),
Bott-Virasoro central extensions and super-symmetry groups.

\paragraph{Scattering.} It might be interesting to explore
the relation of the singular solution momentum map
\textup{(\ref{m-ansatz-intro})} to integrability and scattering
data. For example, see \cite{Va2003} for an interesting
discussion of the Poisson bracket for the scattering data of CH
in 1D. This turns out to be the Atiyah-Hitchin bracket, which is
also related to the Toda lattice, and this fascinating
observation leads to an infinite-dimensional version of Jacobi
elliptic coordinates. 

\paragraph{Other Issues.} Of course there are many other issues
remaining to explore that are suggested by the above setting,
such as convexity of the momentum map, its extension to
Riemannian manifolds, etc. We shall, however,  leave these issues
for other publications and other researchers.

\subsection*{Acknowledgements}
We are very grateful to Alan Weinstein for his collaboration,
help and inspiring discussions over the years. We also thank
Martin Staley for letting us illustrate some important points
using his computations.
\medskip

DDH is grateful for support by US DOE, under contract
W-7405-ENG-36 for Los Alamos National Laboratory, and Office of
Science ASCAR/AMS/MICS. The research of JEM was partially
supported by the California Institute of Technology, the
National Science Foundation through the NSF Grant DMS-0204474
and by the Air Force contract F49620-02-1-0176.

\end{document}